\documentclass[aps,twocolumn,nofootinbib,endfloats]{revtex4-1}

\usepackage{graphicx}
\usepackage{dcolumn}
\usepackage{bm}

\usepackage{hyperref}
\hypersetup{
    colorlinks=true,
    linkcolor=blue,
    citecolor=blue,
    urlcolor=blue
}

\usepackage{graphicx}

\usepackage{titlesec}
\usepackage{amssymb,amsfonts}
\usepackage{epsfig}
\usepackage{epstopdf}
\usepackage[all]{xy}
\usepackage{amsthm}
\usepackage{dcolumn}
\usepackage{hyperref}
\usepackage{url}
\usepackage{dsfont}
\usepackage{slashed}
\usepackage{mathrsfs}

\usepackage{bbm}
\usepackage{tikz}
\usepackage{physics}
\usepackage{enumitem}
\usepackage{soul}
\usepackage{float}

\newcommand{\be}{\begin{equation}}
\newcommand{\ee}{\end{equation}}
\newcommand{\bea}{\begin{eqnarray}}

\newcommand{\eea}{\end{eqnarray}}

\def\inbar{\,\vrule height1.5ex width.4pt depth0pt}
\def\IR{\relax{\rm I\kern-.18em R}}
\def\IC{\relax\hbox{$\inbar\kern-.3em{\rm C}$}}

\begin{document}

\title{Matter-antimatter (a)symmetry in de Sitter Universe}

\author{Jean-Pierre Gazeau$^{1}$\footnote{gazeau@apc.in2p3.fr}}

\author{Hamed Pejhan$^{2}$\footnote{pejhan@math.bas.bg}}

\affiliation{$^1$Universit\'e de Paris, CNRS, Astroparticule et Cosmologie, F-75013 Paris, France\\email: gazeau@apc.in2p3.fr}

\affiliation{$^2$Institute of Mathematics and Informatics, Bulgarian Academy of Sciences, Acad. G. Bonchev Str. Bl. 8, 1113, Sofia, Bulgaria\\email: pejhan@math.bas.bg}

\date{\today}

\begin{abstract}
We investigate the matter-antimatter properties of elementary systems, modeled as free quantum fields, within the global structure of de Sitter spacetime. By leveraging the distinctive causal and analytic properties of de Sitter spacetime, we propose that matter-antimatter asymmetry could emerge as an observer-dependent effect shaped by time orientation within a local causal patch, rather than as a fundamental property of de Sitter Universe itself. This kinematic perspective complements, rather than replaces, standard dynamical processes — such as baryon number violation, $\texttt{CP}$ violation, and nonequilibrium processes — that fulfill Sakharov’s criteria. Within this framework, the limited presence of antimatter in our predominantly matter-filled Universe, specifically within the causal patch of de Sitter spacetime under consideration, may arise from these mechanisms, though through pathways distinct from conventional interpretations.
\end{abstract}

\maketitle

\setcounter{equation}{0}
$\texttt{CP}$ asymmetry — the violation of combined charge conjugation ($\texttt{C}$) and parity ($\texttt{P}$) symmetries — constitutes the primary nontrivial distinction observed to date between matter and antimatter. An antiparticle ($\texttt{CP}$) behaves as the time-reversed ($\texttt{T}$) counterpart of a particle, following the Feynman-Stueckelberg interpretation.

In this Letter, building on this foundational perspective (specifically, the Feynman-Stueckelberg interpretation), we investigate the matter-antimatter characteristics of elementary systems, modeled as free quantum fields, within the global structure of $d$-dimensional de Sitter (dS$^{}_d$) spacetime. dS$^{}_d$ spacetime, the maximally symmetric solution to the vacuum Einstein equations with a positive cosmological constant, is characterized by accelerated expansion and serves as a fundamental model for exploring the dynamics of the contemporary Universe.

\section{Matter-antimatter asymmetry?}
Topologically, dS$^{}_d$ is homeomorphic to $\mathbb{R}^1 \times \mathbb{S}^{d-1}$. Geometrically, the most illustrative representation of dS$^{}_d$ is as a one-sheeted hyperboloid embedded in a $(1+d)$-dimensional Minkowski spacetime $\mathbb{R}^{1+d}$:
\begin{align}
    {\text{dS}^{}_d} = \Big\{ x \in\mathbb{R}^{1+d} \;;\; (x)^2 = x\cdot x = \eta^{}_{\alpha\beta} x^\alpha x^\beta = -H^{-2} \Big\}\,,\nonumber
\end{align}
where $\alpha,\beta = 0,1,\ldots,d$, $\eta^{}_{\alpha\beta} = \mbox{diag}(1,-1,\ldots,-1)$, and $H$ is a positive constant. In the real dS$^{}_{d=4}$ case, $H$ corresponds to the Hubble constant.\footnote{For clarity, we use natural units throughout this Letter, setting the fundamental constants $c$ and $\hbar$ to unity.} 

The dS$^{}_d$ (relativity) group is SO$_0(1,d)$, the connected subgroup of O$(1,d)$. The corresponding Lie algebra is represented by the linear span of the Killing vectors $K_{\alpha\beta} = x_\alpha\partial_\beta - x_\beta\partial_\alpha$. However, none of these vectors is globally timelike, meaning that \underline{neither time nor energy} \underline{can be defined globally in dS$^{}_d$ spacetime \cite{dSBook}}.

This observation leads to our first key insight:\\
\emph{\textbf{Remark 1:} An elementary system, defined as a free quantum field, resides within the global structure of dS$^{}_d$ spacetime. The absence of a globally timelike Killing vector in dS$^{}_d$ spacetime implies that ${\texttt{T}}$ symmetry cannot be globally defined in the conventional sense of inverting a universal time coordinate. Consequently, at least in the sense given by Feynman-Stueckelberg, we cannot unambiguously distinguish between the matter and antimatter descriptions of the system in the global structure of dS$^{}_d$ spacetime. Instead, ${\texttt{T}}$ symmetry can be understood as a local symmetry, applicable in regions where a well-defined notion of time exists, as we elaborate on now.}

\begin{figure}[H]
    \begin{center}
    \includegraphics[height=.38\textheight]{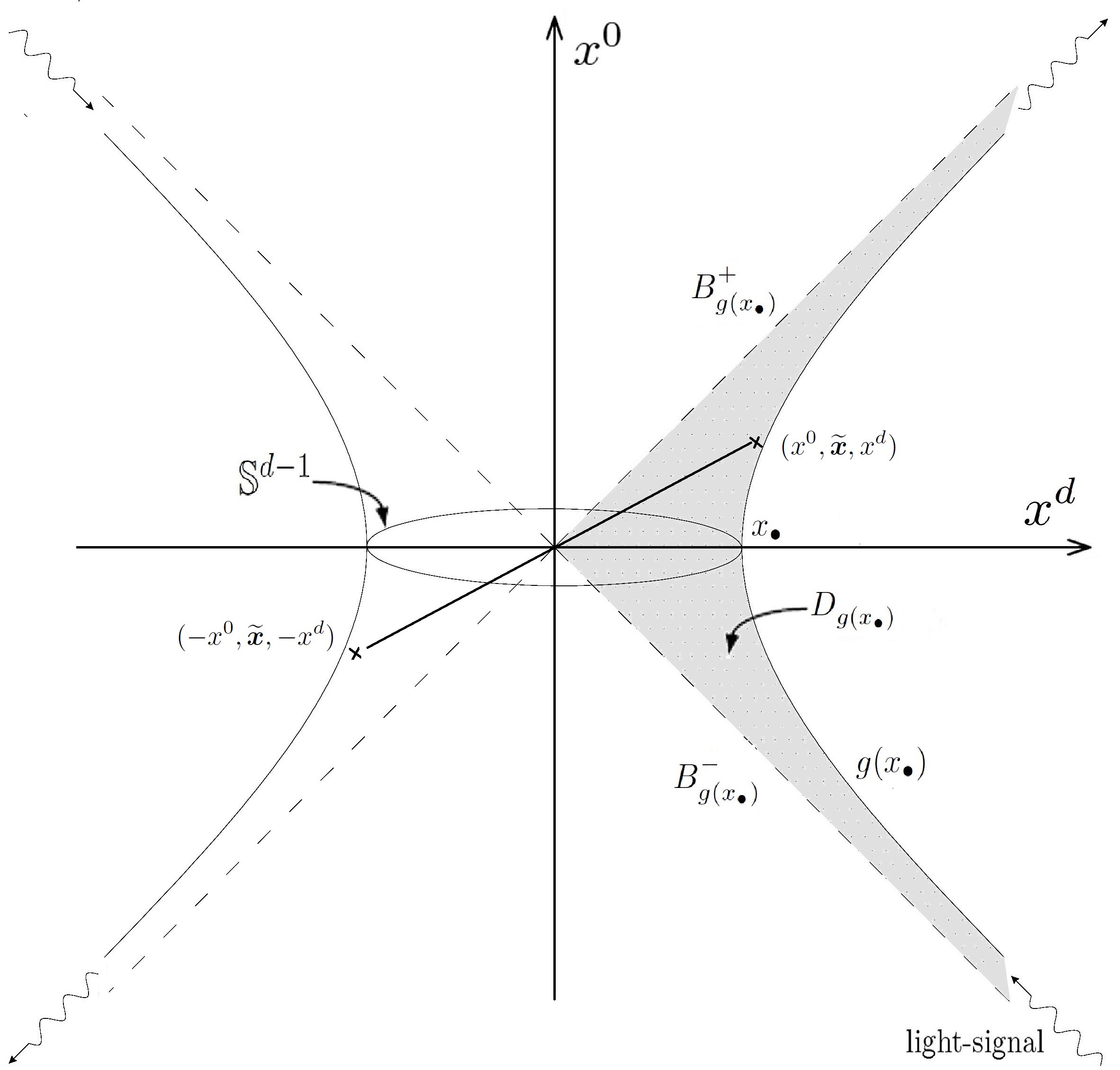}
    \end{center}
    \caption{The static patch of dS$^{}_d$, linked with an observer traveling along the geodesic $g(x_\bullet)$.}
    \label{FIG. KMS}
\end{figure}

Following the approach outlined in Ref. \cite{GibbonsHawking}, let us adopt the viewpoint of a local observer (on dS$^{}_d$) in motion along the geodesic $g({x_\bullet})$ that traverses through the point ${x_\bullet}=(0,\ldots,0, x^d = H^{-1})$ (see FIG. \ref{FIG. KMS}):
\begin{align}\label{geodesic}
    g({x_\bullet}) =\, \Big\{ x=x(t) \; ; \;  x^0 =&\, H^{-1}\sinh{Ht},\nonumber\\ 
    {\boldsymbol{x}} =&\, (x^1,\ldots,x^{d-1})=0, \nonumber\\
    x^{d} =&\, H^{-1}\cosh{Ht}, \quad t\in\mathbb{R} \Big\}\,.
\end{align}
Note that: (i) The chosen point $x_\bullet$ is located in the $(x^0, x^{d})$-plane. (ii) The choice of this point is entirely arbitrary, given the SO$_0(1, d)$ symmetry of dS$^{}_d$.

The set of all ``events'' in ${\text{dS}^{}_d}$ that can be linked to the observer via the reception and emission of light signals is given by $D^{}_{g({x_\bullet})} = \big\{ x \in {\text{dS}^{}_d} \; ; \; x^{d} > |x^0| \big\}$. This domain is delineated by two distinct boundaries ${B}^{\pm}_{g({x_\bullet})} = \big\{ x \in {\text{dS}^{}_d} \; ; \; x^0 = \pm x^{d}, \; x^{d}>0 \big\}$. These boundaries are respectively termed the ``future horizon'' and ``past horizon'' of the observer following the geodesic ${g({x_\bullet})}$.

The parameter $t$ in expression \eqref{geodesic} signifies the proper time of the observer moving along the geodesic ${g({x_\bullet})}$. This enables us to designate the ``time-translation group linked with ${g({x_\bullet})}$'' as the one-parameter subgroup ${\mathfrak{T}}^{}_{g({x_\bullet})}$ (isomorphic to $\mathrm{SO}_0(1,1)$) of the dS$^{}_d$ group $\mathrm{SO}_0(1,d)$. The ${\mathfrak{T}}^{}_{g({x_\bullet})}$'s transformations involve hyperbolic rotations parallel to the $(x^0, x^{d})$-plane. To characterize how these transformations operate on the domain $D^{}_{g({x_\bullet})}$, let $x=x(t, \widetilde{\boldsymbol{x}})$ represent an arbitrary point within this domain:
\begin{eqnarray}\label{coordinate KMS}
    x(t,{\widetilde{\boldsymbol{x}}}) =
    \left \{ \begin{array}{rl} x^0 &= \sqrt{H^{-2}-(\widetilde{\boldsymbol{x}})^2} \; \sinh{Ht}\,, \vspace{2mm}\\
    \vspace{2mm} \widetilde{\boldsymbol{x}} &= (x^1,\ldots,x^{d-1})\,,\\
    \vspace{2mm} x^{d} &= \sqrt{H^{-2}-(\widetilde{\boldsymbol{x}})^2} \; \cosh{Ht}\,, \quad t\in\mathbb{R}\,, \end{array}\right.
\end{eqnarray}
while $(\widetilde{\boldsymbol{x}})^2=({x}^1)^2+\ldots+({x}^{d-1})^2 < H^{-2}$. The operation of ${\mathfrak{T}}^{}_{g({x_\bullet})}$ on $x(t,\widetilde{\boldsymbol{x}})$ is expressed as follows:
\begin{align}
    {\mathfrak{T}}^{}_{g({x_\bullet})}({\tau}) \diamond x(t,\widetilde{\boldsymbol{x}}) = x(t+\tau,\widetilde{\boldsymbol{x}}) = x^\tau_{}\,, \quad \tau\in\mathbb{R} \,.\nonumber
\end{align}
It defines a group of isometric automorphisms within the domain $D^{}_{g({x_\bullet})}$. The associated orbits $g^{}_{\widetilde{\boldsymbol{x}}}({x_\bullet})$ depict separate hyperbolic branches in $D^{}_{g({x_\bullet})}$. These hyperbolas lie in two-dimensional plane sections that are parallel to the $(x^0, x^{d})$-plane. Within these orbits, the singular one that defines a geodesic of dS$^{}_d$ is $g({x_\bullet}) = g^{}_{\widetilde{\boldsymbol{x}}=\boldsymbol{0}}({x_\bullet})$. Hence, \underline{the} \underline{time-translation group ${\mathfrak{T}}^{}_{g({x_\bullet})}$ is particularly relevant for} \underline{observers in motion on or near $g({x_\bullet})$}, provided that their proximity is significantly smaller than the dS$^{}_d$ radius of curvature.

This brings us to the second remark:\\
\emph{\textbf{Remark 2:} When a local observer fixes a time direction (proper time $t$), this act breaks the $\texttt{T}$ symmetry within the observable region of dS$^{}_d$ spacetime (i.e., $D^{}_{g({x_\bullet})}$). Consequently, at a given point $(x^0,\widetilde{\boldsymbol{x}},x^{d})$ within this causal domain accessible to the observer, the elementary system is identified as matter. This local identification reflects the observer's choice of time direction (in the sense given by Feynman-Stueckelberg) and aligns with the observable properties of the quantum field in this region.}

To support future reasoning, it is essential to add that:\\
\emph{\textbf{Remark 3-1:} By establishing a preferred direction of time, the observer also \underline{implicitly} defines the reversed direction of time in the corresponding mirror region $D^{}_{g(-{x_\bullet})} = \big\{ x \in {\text{dS}^{}_d} \; ; \; -x^{d} < -|x^0| \big\}$. Strictly speaking, the intrinsic time variable relevant to an observer moving along the mirror geodesic $g(-{x_\bullet})$, within $D^{}_{g(-{x_\bullet})}$, corresponds to $-t$. This arises from the fact that the associated time-translation group ${\mathfrak{T}}^{}_{g(-x_\bullet)}$ validates the relation ${\mathfrak{T}}^{}_{g(-x_\bullet)} (\tau) = {\mathfrak{T}}^{}_{g(x_\bullet)} (-\tau)$. Note that ${\mathfrak{T}}^{}_{g(-x_\bullet)}$ is derived from ${\mathfrak{T}}^{}_{g(x_\bullet)}$, for instance, through conjugation in the form ${\mathfrak{T}}^{}_{g(-x_\bullet)} = R\,{\mathfrak{T}}^{}_{g(x_\bullet)}\, R^{-1}$, where $R$ represents a rotation of angle $\pi$ in a plane orthogonal to the $x^0$-axis.\footnote{This characteristic serves as yet another indication of the fact that dS$^{}_d$ spacetime lacks a globally timelike Killing vector.}}

\emph{\textbf{Remark 3-2:} The two corresponding mirror regions, $D^{}_{g({x_\bullet})}$ and $D^{}_{g(-{x_\bullet})}$, are causally disconnected. Therefore, observations made in one region have no causal influence on the mirror region.}

Another foundational element of our reasoning lies in the imperative of analyticity within the complexified dS$^{}_d$ manifold (denoted as dS$^{(\mathbb{C})}_d$):
\begin{align}
    &\text{dS}_d^{(\mathbb{C})} = \Big\{z =  x+ \mathrm{i} y \in {\mathbb{C}}^{1+d} \;;\; (z)^2= \eta^{}_{\alpha\beta} z^\alpha z^\beta = -H^{-2} \nonumber\\
    &\qquad\mbox{or equivalently} \;;\; (x)^2 - (y)^2 = -H^{-2}, \; x\cdot y=0 \Big\}\,.\nonumber
\end{align}
This analyticity requirement is crucial for the quantum field theory (QFT) describing elementary systems within the global structure of dS$^{}_d$. Expanding beyond the familiar flat Minkowski spacetime to dS$^{}_d$ introduces a fundamental challenge to the QFT description of elementary systems. The challenge lies in the absence of a well-defined spectral condition \cite{Streater}. Regardless of the chosen QFT approach to field quantization on dS$^{}_d$ spacetime, formulating requirements for locality (microcausality) and covariance is a relatively straightforward task. However, establishing any condition on the spectrum of the ``energy'' operator proves to be a formidable challenge. As previously noted, in dS$^{}_d$ spacetime, granted that no globally timelike Killing vector exists, neither time nor energy can be globally defined. This inherent ambiguity results in the emergence of numerous inequivalent QFTs (vacuum states) for a single field model on dS$^{}_d$ spacetime. Each of these theories is typically linked to a specific choice of the time coordinate, leading to associated frequency splitting. 

In the Minkowskian framework, a notable observation highlights a direct link between the characteristics of analytic continuation in complexified spacetime and the spectral property of the model under consideration. Specifically, in the context of $(1+d)$-dimensional Minkowski spacetime $\mathbb{R}^{1+d}$, the spectral property can be expressed as follows: for each $n>1$, the Wightman $n$-point function corresponds to the boundary value, in the distribution sense, of a function that is holomorphic within the tube $T^{+(n)}$ defined as:
\begin{align}
    T^{+(n)} = \Big\{ ( z^{}_1, \ldots, z^{}_n ) \in \mathbb{C}^{n(1+d)} \;;\;  \Im(z^{}_{j+1} - z^{}_j) \in {{V}}^+ \Big\}\,,\nonumber
\end{align}
where $1\leqslant j \leqslant n-1$ and ${V}^\pm = \big\{y \in\mathbb{R}^{1+d} \;;\; (y)^2 > 0,\; y^0 \gtrless 0 \big\}$.

In the dS$^{}_d$ context (embedded in $\mathbb{R}^{1+d}$), a natural alternative to this property is to propose that the Wightman $n$-point function ${\cal{W}}_n(x_1,\ldots,x_n)$ acts as the boundary value, in the distributional sense, for a function that is holomorphic within ${\cal{T}}^{+(n)} = T^{+(n)} \cap \big[\text{dS}_d^{(\mathbb{C})}\big]^n$. Studies by Bros et al. \cite{GazeauPRL, Bros2pointfunc} showed that ${\cal{T}}^{+(n)}$ forms a domain and a tuboid — bounded by reals — in a manner that maintains the relevance of the concept of the ``distribution boundary value of a holomorphic function from this domain''. Consequently, it becomes feasible to impose \underline{weak spectral condition}: 
\begin{itemize}
    \item{For each $n>1$, the Wightman $n$-point function ${\cal{W}}_n(x_1,\ldots,x_n)$ is derived from the distributional boundary value of a function $W_n(z_1,\ldots,z_n)$, which is holomorphic within the domain ${\cal{T}}^{+(n)}$.}
\end{itemize}
Focusing on a generalized free dS$^{}_d$ field, the assertion of normal analyticity can be explicitly expressed as: 
\begin{itemize}
    \item{The corresponding two-point function ${\cal{W}}_2(x_1,x_2)$ arises as the distributional boundary value of a function $W_2(z_1,z_2)$ that is analytically defined within the tuboid domain ${\cal{T}}^{+(2)}_{}$ given by ${\cal{T}}^{+(2)}_{} = \big\{ (z_1,z_2) \; ; \; z_1\in {\cal{T}}_{}^-, \; z_2 \in {\cal{T}}_{}^+ \big\}$, where ${\cal{T}}^\pm = \big\{ \mathbb{R}^{d+1} + \mathrm{i} {V}^\pm \big\} \cap \text{dS}_d^{(\mathbb{C})}$.}
\end{itemize}
Note that ${\cal{T}}^\pm$ are referred to as the forward and backward tubes of $\text{dS}_d^{(\mathbb{C})}$, respectively.

This elegant resolution effectively circumvents all ambiguities in dS$^{}_d$ QFTs, which arise from the absence of a genuine spectral condition in dS$^{}_d$ spacetime \cite{GazeauPRL, Bros2pointfunc}. It leads to the derivation of ``vacua'' that, despite exhibiting thermal properties, serve as precise analogs of Minkowski vacuum representations. The latter emerges as the limit of the former when the curvature tends to zero.

Now, let us examine the depicted scenario for the local observer traveling along the geodesic $g({x_\bullet})$ by acknowledging the necessity of analyticity within the complexified dS$^{}_d$ manifold (dS$^{(\mathbb{C})}_d$). The complex orbits of ${\mathfrak{T}}^{}_{g({x_\bullet})}$ are denoted as:
\begin{align}\label{EEE}
    g^{(\mathbb{C})}_{\widetilde{\boldsymbol{x}}}({x_\bullet}) = \Big\{ z^\tau_{} = z(t+\tau,\widetilde{\boldsymbol{x}}),\; \tau\in\mathbb{C} \Big\}\,.
\end{align}
Note that all nonreal points linked to the complex orbits $g^{(\mathbb{C})}_{\widetilde{\boldsymbol{x}}}({x_\bullet})$ reside in ${\cal{T}}^\pm$, the very domains of analyticity required for the QFT description of elementary systems in the global structure of dS$^{}_d$ spacetime. In this context, due to the analytic nature of dS$^{}_d$ QFTs, a notable connection arises between points/``events'' $x=(x^0,\widetilde{\boldsymbol{x}},x^{d})$ in the domain $D^{}_{g({x_\bullet})}$ and their corresponding points/``events'' $(-x^0,\widetilde{\boldsymbol{x}},-x^{d})$ in the mirror region $D^{}_{g(-{x_\bullet})}$, as shown in FIG. \ref{FIG. KMS}; to see the point, with the coordinate system \eqref{coordinate KMS} in consideration, setting $\Im(\tau)=\pi/H$ in \eqref{EEE} suffices.

This key property underpins the basis of our concluding remark:\\
\emph{\textbf{Remark 4-1:} Returning to the elementary system discussed in Remarks 1 and 2, modeled as a free quantum field within the global structure of dS$^{}_d$ spacetime, we consider the perspective of the observer in Remark 2. When this observer identifies a matter realization of the system at a given point $(x^0, \widetilde{\boldsymbol{x}}, x^d) \in D^{}_{g({x_\bullet})}$, the analytic properties of dS$^{}_d$ QFTs immediately imply an equivalent description of the system at the causally disconnected mirror point $(-x^0, \widetilde{\boldsymbol{x}}, -x^d) \in D^{}_{g(-{x_\bullet})}$, but with the time orientation reversed (see Remark 3). Under the Feynman-Stueckelberg interpretation, this mirrored realization represents the antimatter counterpart of the original matter realization at $(x^0, \widetilde{\boldsymbol{x}}, x^d)$.\footnote{It is crucial to underline that this observation aligns with the fact that the opposing static patches of dS$^{}_d$ spacetime are linked by a bifurcate Killing horizon, similar to the two opposing Rindler wedges in Minkowski spacetime \cite{cc}. Globally regular quantum states in dS$^{}_d$ spacetime, therefore, exhibit correlations between these opposing static patches. In the case of a charged field, these correlations involve charge conjugation, much like the situation in Rindler spacetime, as discussed in Ref. \cite{qq}.} Thus, as far as free quantum fields are concerned, we argue that while a local observer perceives a matter-antimatter asymmetry, this asymmetry is a consequence of their limited causal perspective, rather than an inherent feature of the global structure of dS$^{}_d$ spacetime.}

\emph{\textbf{Remark 4-2:} It is crucial to underline that the kinematical perspective introduced in this Letter is not intended as a replacement for the established dynamical processes outlined by Sakharov's criteria — specifically, baryon number violation, $\texttt{CP}$ violation, and deviation from thermal equilibrium. Instead, it serves as a complementary interpretation. We propose that traces of antimatter observed within our predominantly matter-filled Universe (or, more precisely, within the causal patch of the dS$^{}_d$ Universe considered here) might indeed stem from these dynamical processes but through mechanisms distinct from the conventional view.}

\emph{Although previous studies have shown that $\texttt{CP}$ asymmetry in particle physics alone cannot fully explain the pronounced matter-antimatter imbalance favoring matter, we suggest that, under certain conditions favoring antimatter, this $\texttt{CP}$ asymmetry could account for the limited antimatter presence observed within our matter-dominated Universe (again, specifically within the dS$^{}_d$ causal patch under consideration).\footnote{So far, the best-documented examples of $\texttt{CP}$ violation with possible antimatter preference occur in the $B$-meson system, where direct  $\texttt{CP}$ violation in decay processes sometimes favors antimatter states. Additionally, if  $\texttt{CP}$ violation is confirmed in neutrino oscillations, it could point to a fundamental asymmetry that favors antimatter (see, for instance, Refs. \cite{111,222,333}).} This interpretation, framed within our kinematical approach, thus explores how $\texttt{CP}$-violating processes might account for this subtle yet non-negligible antimatter component, despite the overwhelming dominance of matter.}

In conclusion, we note that the dS$^{}_{d=4}$ Universe presented here is an idealized model for real-world cosmology. Although the cosmological constant aligns with supernova data, suggesting a dS$^{}_{d=4}$-like expansion, potential indications of dynamical dark energy, as hinted by DESI data, could challenge this exact dS$^{}_{d=4}$ form. Furthermore, while the early Universe may have undergone accelerated expansion (e.g., during inflation), this expansion likely does not strictly conform to the dS$^{}_{d=4}$ model.

To address these challenges, it is crucial to extend our approach — particularly the key analytic insights leading to Remark 4, which, outside of the maximally symmetric cases (Minkowski, dS$^{}_d$, and Anti-dS$^{}_d$), remains largely unexplored — toward less idealized cosmological models, such as the Friedmann-Robertson-Walker spacetime, which asymptotically resembles dS$^{}_d$. This extension would provide more robust and comprehensive results, offering insights better suited to broader and more realistic cosmological applications.

\section{Discussion:\\Applicability of the Feynman-Stueckelberg interpretation in stationary vs. nonstationary spacetimes}
A key consideration in the traditional application of the Feynman-Stueckelberg interpretation for distinguishing particles from antiparticles is that it typically assumes a stationary spacetime. In such stationary spacetimes, globally well-defined concepts of time and energy enable an unambiguous identification of particles and antiparticles based on consistent energy flow and time direction.

In contrast, our work extends the Feynman-Stueckelberg interpretation to dS$^{}_d$ spacetime, which, as we note, is not globally stationary. dS$^{}_d$ spacetime lacks a globally timelike Killing vector, meaning there is no universal definition of time or energy across the entire spacetime. Instead, we focus on the causal patches accessible to local observers within dS$^{}_d$ spacetime, where a well-defined notion of time exists. 

In this context, by considering the analytic continuation in the complexified dS$^{}_d$ spacetime (denoted dS$^{(\mathbb{C})}_d$), we exploit its inherent symmetry and analyticity to establish a ``mirror'' relationship between causally disconnected regions. This structure allows us to interpret the matter manifestation of an elementary system (a globally defined free quantum field) at a given point in one causal patch as having a corresponding antimatter manifestation at the mirror point in a causally disconnected region, where time flows in the opposite direction. The mirror point, mathematically defined by extending coordinates into the complex plane, directly represents the antimatter counterpart of the original matter realization.

Thus, our approach for identifying antiparticles in dS$^{}_d$ spacetime does not rely on global stationarity. Instead, it leverages the unique analytic continuation and causal structure of dS$^{}_d$, enabling a local, observer-dependent adaptation of the Feynman-Stueckelberg interpretation. Here, particle-antiparticle distinctions are defined within individual causal patches, rather than globally.

In summary, our analysis modifies the Feynman-Stueckelberg framework to accommodate the nonstationary nature of dS$^{}_d$ spacetime by making use of its causal and analytic structure. This yields a region-specific approach to particle-antiparticle identification, which does not depend on global stationarity.

\section*{Acknowledgment} 
Hamed Pejhan is supported by the Bulgarian Ministry of Education and Science, Scientific Programme `Enhancing the Research Capacity in Mathematical Sciences (PIKOM)', No. DO1-67/05.05.2022.


\end{document}